\documentclass[letterpaper]{article} 
\usepackage{aaai24}  
\usepackage{times}  
\usepackage{helvet}  
\usepackage{courier}  
\usepackage[hyphens]{url}  
\usepackage{graphicx} 
\usepackage{natbib}  
\usepackage{caption} 
\usepackage{algorithm}
\usepackage{algorithmic}
\usepackage{amsfonts}

\usepackage{newfloat}
\usepackage{listings}
\DeclareCaptionStyle{ruled}{labelfont=normalfont,labelsep=colon,strut=off} 
\floatstyle{ruled}
\newfloat{listing}{tb}{lst}{}
\floatname{listing}{Listing}
\title{Temporal Fairness in Multiwinner Voting}
\author {
    Edith Elkind\textsuperscript{\rm 1,2},
    Svetlana Obraztsova\textsuperscript{\rm 3},
    Nicholas Teh\textsuperscript{\rm 1}
}
\affiliations {
    \textsuperscript{\rm 1}  University of Oxford, United Kingdom\\
    \textsuperscript{\rm 2} Alan Turing Institute, United Kingdom\\
    \textsuperscript{\rm 3} Carleton University, Canada\\
    elkind@cs.ox.ac.uk, svetlanaobraztsova@cunet.carleton.ca, nicholas.teh@cs.ox.ac.uk
}

\usepackage{bibentry}

\begin{document}

\maketitle

\begin{abstract}
Multiwinner voting captures a wide variety of settings, 
from parliamentary elections in democratic systems to product placement in online shopping platforms. There is a large body of work dealing with axiomatic characterizations, computational complexity, and algorithmic analysis of multiwinner voting rules. Although many challenges remain, significant progress has been made in showing existence of fair and representative outcomes as well as efficient algorithmic solutions for many commonly studied settings. However, much of this work focuses on single-shot elections, even though in numerous real-world settings elections are held periodically and repeatedly.
Hence, it is imperative to extend the study of multiwinner voting to temporal settings.
Recently, there have been several efforts to address this challenge.
However, these works are difficult to compare, as they model multi-period voting in very different ways.
We propose a unified framework for studying temporal fairness in this domain, drawing connections with various existing bodies of work, and consolidating them within a general framework.
We also identify gaps in existing literature, outline multiple opportunities for future work, and put forward a vision for the future of multiwinner voting in temporal settings.
\end{abstract}

\section{Introduction}
Elections for the AAAI Executive Council take place annually, with one-third 
of the positions being up for election each time. Artificial Intelligence research 
encompasses a broad variety of topics, and it is important that the council has representatives
from all major subfields. Moreover, the community is very international, 
and the council should represent the interests of members around the world, 
so there is a desire for \emph{proportionate representation} in terms of geographical regions. 
Diversity in terms of gender and academic seniority is
another key desideratum. 

There are several multiwinner voting rules that provide fair representation 
guarantees~\cite{aziz2017jr,peters2020proportionality}. However, existing notions of fairness 
for multiwinner voting do not fully capture the complexity of our setting: it may be the case that
a voter is not represented by any of the candidates elected in the current round, yet
she is satisfied with the overall composition of the council because she is
well-represented by some of the council members elected in the previous rounds.
Also, given that the council is renewed regularly, perhaps a group of voters that
is too small to be assured permanent representation on the council can be guaranteed
a spot every few rounds? 

In a somewhat different spirit, consider proof-of-stake blockchain protocols where 
a primary concern is  the 
``rich gets richer effect''~\cite{fanti2019cryptocompound,huang2021richgetricher}.
In such scenarios, lotteries for a single-shot interaction provide a form of fairness, as the probability of winning is proportional to the invested effort~\cite{orda2019blockchain,pan2022fairblockchain}. 
However, such mechanisms may fail to maintain fairness \emph{over time} 
when lotteries are held repeatedly over multiple 
rounds~\cite{fanti2019cryptocompound,grossi2022socialchoiceblock}. 
Formalizing fairness concepts and successfully designing mechanisms 
for achieving these in the repeated-interaction setting 
is therefore highly relevant to blockchain systems.

There are further examples of real-world elections where
temporal considerations play an important role, and many 
authors explored multiwinner voting with temporal 
elements. However, this body of work does not yet offer a systematic exploration of
issues associated with voting over time, with each strand of research
considering a slightly different model and a different set of applications. 
Against this background, we propose a unified framework to facilitate a principled study 
of temporal fairness in multiwinner voting, highlighting several key challenges,  
consolidating existing bodies of work, and identifying gaps in the 
existing literature.
We aim to set the groundwork for a more coherent, targeted approach to tackling 
conceptual and algorithmic challenges in temporal multiwinner voting.

Several elements of our framework originate from a larger body of work 
within the social choice literature and beyond.
This includes, in particular, works on
\emph{perpetual voting} \cite{lackner2020perpetual,lackner2023proportionalPV}---and, more broadly, perpetual participatory budgeting \citep{lackner2021longtermpb}--- 
which consider multiple rounds of single-winner voting. 
\citet{lackner2020perpetual} and
\citet{lackner2023proportionalPV}
conduct an axiomatic analysis of perpetual voting rules 
(temporal extensions of traditional voting rules) with respect to fairness across time. Relatedly, \citet{harrenstein2022election} argue
that several common 
sequential election mechanisms may be detrimental to welfare.
Another relevant strand of work is that on conference scheduling \cite{patro2022virtualconf}, 
slot assignment \cite{elkind2022temporalslot}, 
and line-up elections \cite{boehmer2020lineupelections}, 
where the goal is to output a sequence of non-repeating winners, 
whilst maintaining fairness to voters across the entire time horizon. 
Sequential committee election 
models 
consider settings where an entire committee is elected in each round, and impose constraints 
on the extent a committee can change, whilst ensuring candidates continue 
to have support from the electorate \citep{bredereck2020successivecommittee,bredereck2022committeechange,deltl2023seqcommittee}. 
Other models in the social choice literature that include temporal elements include sequential decision-making \citep{chandak2023sequential,kahana2023seqcollectiveleximin},
public decision-making \citep{brandt2016handbook,conitzer2017fairpublic,fain2018publicgoods,skowron2022proppublic,lackner2023freeriding}, repeated matching \citep{gollapudi2020repeatedmatching,trabelsi2023matchings,caragiannis2023repeatedmatching}, resource allocation over time \citep{bampis2018fairtime,igarashi2023repeated}, online committee selection \citep{do2022onlinecommittee}, dynamic social choice \citep{parkes2013dynamicsocialchoice,freeman2017dynamicsocialchoice}, and temporal liquid democracy \citep{markakis2023temporalld}.

As illustrated by the many examples mentioned above, 
maintaining welfare and fairness guarantees over time is relevant to many domains. 
Research into temporal multiwinner elections would thus allow us 
to capture many practical scenarios, discover notions of representation
that are appropriate for temporal settings, and offer insights 
into the possibility of achieving desirable goals by society
in a computationally efficient manner.

The remainder of the paper is structured as follows. First, 
we introduce the framework. We then proceed to look at each element of the model, 
fleshing out various options available, consolidating existing work and positioning it with regard to these options, and pointing out gaps or ideas for expansion. 
Finally, to set the foundations for further study,
we formulate several research challenges and potential directions 
to push this field forward, and highlight the immense potential of research in this area.

\section{The Temporal Multiwinner Voting Framework} \label{sec:framework}
The basic components of our framework are a set $N = \{1,\dots,n\}$ of $n$ \emph{agents}, 
a set $P = \{p_1,\dots,p_m\}$ of $m$ distinct {\em projects}, or \emph{candidates}, 
and a set $[\ell] = \{1,\dots, \ell\}$ of $\ell$ \emph{timesteps}.
For each timestep $r \in [\ell]$, each agent $i \in N$ has a {\em preference} $s_{i,r}$. 
This can come in various forms (see the Preferences section for a discussion).
We write $\mathbf{s}_i = (s_{i,1},s_{i,2},\dots, s_{i,\ell})$ 
and refer to $\mathbf{s}_i$ as $i$'s \emph{temporal preference}.
The goal is to select an \emph{outcome} $\mathbf{o} = (o_1,\dots,o_\ell)$, 
which is a sequence of $\ell$ sets of candidates such that for every $r \in [\ell]$, 
the set of candidates $o_r \subseteq P$ is chosen at timestep $r$.

Next, we discuss various elements of the framework, contrasting several options 
within each element, and positioning the existing work in the 
computational social choice literature with respect to these options.
We also outline several directions for future work.

\section{Outcomes} \label{sec:outcomes}
The first parameter we explore is what is considered a permissible outcome at each timestep. 
In particular, we distinguish between selecting one candidate at a time
and choosing multiple candidates at each timestep.
We also consider constraints that can be imposed on the outcome vector.

\subsection{Structure}
In a single-shot multiwinner election, the outcome is a single set of candidates.
There are two possible generalizations to the temporal setting.
The first (and one that is more common in the literature) is having a single winning candidate chosen 
at each timestep (i.e., $o_r\in P$ for each $r \in [\ell]$). While one can view
this variant of the model as a temporal extension of single-winner elections, 
the multiwinner interpretation is justified, too, as one can treat the (multi-)set
$O = \{o_r: r\in [\ell]\}$ as the winning committee and apply fairness concepts
that originate in multiwinner voting literature to the entire set $O$; e.g., 
\citet{bulteau2021jrperpetual} and \citet{page2021electing} reason about justified representation 
provided by $O$.
This model is considered in numerous existing works, 
including scheduling problems \citep{elkind2022temporalslot,patro2022virtualconf}, 
the perpetual voting model \cite{lackner2020perpetual,lackner2023proportionalPV}, 
and in public decision-making~\cite{conitzer2017fairpublic,fain2018publicgoods}.
The second model assumes that the goal is to select a fixed-size 
set of winning candidates (i.e., an entire committee) 
at each timestep. This approach is taken, e.g.,  by \citet{bredereck2020successivecommittee}, \citet{bredereck2022committeechange}, and \citet{deltl2023seqcommittee} in their study of 
sequential committee 
elections.

\subsection{Feasibility Constraints}
In the standard multiwinner voting setting, it is typically assumed that there is a parameter
$k$ such that every subset of $P$ of size $k$ is a feasible outcome (we note, however, 
that there is also work on single-shot multiwinner voting with 
constraints on feasible committees~\cite{yang2018mwvrestrictions}). 
However, in temporal settings it may be the case that at each timestep $r$ 
only a subset of candidates $P_r\subseteq P$ is available, and the winning
candidate(s) have to be chosen from this set. The simplest variant of this model
is where the set $P_r$ is given in advance and is independent of committees selected
at steps $r'$ with $r'\neq r$. This model is relevant if changes in $P_r$ are caused
by external constraints in candidate availability (i.e., perhaps candidate X does not want 
to run for the AAAI executive council this year due to heavy administrative load in their department,
but will be happy to serve in the future).
In the context of public decision-making 
\cite{conitzer2017fairpublic,fain2018publicgoods,skowron2022proppublic}, such 
constraints can be 
influenced by the suitability of a project being implemented at a particular timestep due to manpower or geographical constraints~\cite{lodi2022dynamichealthcare}.

However, it may also be the case that a decision at timestep $r$ constrains the options
available at timestep $r'$: e.g., in fair scheduling problems it is common to assume
that each project in $P$ can be implemented at most once, so if we select $p\in P$
at timestep $r$, it is no longer available at 
$r'\neq r$~\cite{elkind2022temporalslot,patro2022virtualconf}. More generally, for
each project $p\in P$ there may be a bound $\alpha_p\in\mathbb N$, indicating 
the maximum number of times that $p$ can be selected. 
One can also consider more sophisticated constraints, where 
at most (at least) a certain
fraction of the winning committee needs to be replaced at each timestep; this approach
is taken by \citet{bredereck2020successivecommittee} and \citet{bredereck2022committeechange}.
There are other constraints of this form that can be found in practice, but, to the best
of our knowledge, have not been modeled in the literature: e.g., an AAAI executive council
member, once elected, remains a member of the winning committee for three timesteps, 
but is then not eligible to participate in the next election (but can run again later on).

\section{Preferences} \label{sec:preferences}
Another important component of our framework is agents' preferences.
The aspects that need to be considered include ballot types (e.g., 
approval, ranked, or cardinal), 
and if  preferences can evolve over time (static or dynamic).

\subsection{Ballot Types} \label{sec:preference_elicitation}
The computational social choice literature typically focuses on three ballot types:
approval ballots~\cite{aziz2015mwvapproval,lackner2022abc}, 
ranked ballots~\cite{faliszewski2017mwv,elkind2017propertiesmwv}, 
and cardinal ballots~\cite{conitzer2017fairpublic,freeman2017dynamicsocialchoice,fain2018publicgoods}.

\subsubsection*{\textbf{Approval ballots}}
Much of the recent work on proportionality in computational social choice
focuses on \emph{approval} ballots \citep{lackner2022abc}: each agent reports which candidates
they like and dislike. Approval preferences are relatively easy to elicit and reason 
about~\cite{kilgour1983av,brams2005av,aragones2011av}, 
yet they can capture a wide variety of scenarios from city budget planning 
to elections for board of trustees. In temporal settings, approval ballots have been considered 
in the context of sequential committee elections 
\cite{bredereck2020successivecommittee,bredereck2022committeechange,deltl2023seqcommittee} and scheduling 
\cite{bulteau2021jrperpetual,elkind2022temporalslot}.

\subsubsection*{\textbf{Ranked ballots}}
Under ranked ballots, each agent reports a \emph{ranking} 
over the candidates at each timestep; voting with ranked ballots
has been extensively studied in single-shot multiwinner elections~\citep{faliszewski2017mwv}, 
with applications ranging from parliamentary elections 
to movie selection~\cite{elkind2017propertiesmwv}.

\subsubsection*{\textbf{Cardinal ballots}}
Cardinal ballots---where each voter explicitly specifies the \emph{utility} they obtain from each candidate---
offer a lot of expressivity, albeit at a higher cost of elicitation.
Such ballots have been studied in settings such as public 
decision-making \cite{conitzer2017fairpublic,freeman2017dynamicsocialchoice,fain2018publicgoods}, portioning \cite{freeman2021truthfulbudget,elkind2023portioning}, and 
line-up elections \cite{boehmer2020lineupelections}.

\subsection{Temporal Evolution}
Next, we consider the temporal evolution of preferences.
Preferences are said to be \emph{static} if they do not change over time, 
and are \emph{dynamic} otherwise.

\subsubsection*{\textbf{Static preferences}}
In this setting, each voter's preferences remain the same across the entire time horizon, 
and temporal considerations arise because of candidate availability issues or constraints
on possible outcomes (refer to the Outcomes section for a discussion). 
For instance, \citet{bulteau2021jrperpetual} consider static preferences
in perpetual voting, and obtain positive algorithmic results for achieving notions of proportionality in this setting.

\subsubsection*{\textbf{Dynamic preferences}}
Dynamic preferences capture the idea that agents' preferences may evolve
with time. Just as in case of outcomes, we distinguish between 
{\em non-adaptive} preferences,
which evolve due to external considerations (e.g., when expressing 
preferences over restaurants for each day of the week, an agent
may prefer a non-meat option on Fridays),
and {\em adaptive} preferences, which change based
on decisions that have been made so far (e.g., an agent
may be unwilling to eat at the same restaurant twice in a row).
For instance, the literature on the building of public projects 
\cite{conitzer2017fairpublic,fain2018publicgoods,skowron2022proppublic} 
and \citet{bulteau2021jrperpetual} work on perpetual voting
 consider dynamic non-adaptive preferences, whereas \citet{parkes2013dynamicsocialchoice} consider dynamic adaptive preferences.

\subsubsection*{\textbf{Intertemporal constraints}}
An important special case that is not captured by the static/dynamic
dichotomy is intertemporal restrictions on voters' ballots:
e.g., in the context of scheduling, an agent may be asked 
to report her ideal schedule \cite{elkind2022temporalslot}.
This means that agents approve exactly one project per timestep, with no repetition of projects---this is equivalent to a constraint imposed across timesteps.

\medskip

We also note that, at least for approval ballots, simple candidate availability constraints
can be incorporated into voters' preferences: e.g., if candidate $p$ is not available
at timestep $r$, we can simply remove $p$ from all voters' ballots at $r$. By doing so, 
one can simplify the description of the input instance. However, this approach is not 
well-suited to more complex feasibility constraints or ranked ballots.

\section{Observability}
The next element we consider is the observability of future timesteps.
The setting is said to be \emph{online} if the preferences of agents at future timesteps are not known in advance, and \emph{offline} otherwise.

\subsubsection*{\textbf{Online settings}}
Online scenarios are most prevalent in the dynamic social choice literature \citep{freeman2017dynamicsocialchoice,freeman2018dynamicproportional}, where candidates are selected sequentially without information on agents' preferences at future timesteps. 
This is reminiscent of the secretary problem, or, more specifically, its $k$-winner variant~\cite{albers2021new}; however, the key difference is that in the standard secretary problem the decision is made by a single stakeholder.
\citet{do2022onlinecommittee} consider a multiple-stakeholder variant of the $k$-secretary problem that specifically addresses the question of whether committees selected online can proportionally represent voters. A similar model is explored by~\citet{israel2021dynamicproprankings}. Another line of work considers settings where voters rather than candidates are the ones that appear online~\cite{oren2014onlinesocialchoice,dey2017propvotestreams}.
Other related online models in social choice include those in fair division (see, e.g., the survey by \citet{aleksandrov2020onlinefairdivision}).

\subsubsection*{\textbf{Offline settings}}
In offline settings, agents' preferences are fully known in advance, prior to the computation of the outcome at each timestep.
Several works on fair scheduling \citep{elkind2022temporalslot,patro2022virtualconf} consider an offline model where preferences are fully available at the onset, and the goal is to study the computational problems associated with finding a desirable outcome or to conduct an axiomatic analysis of the mechanisms for obtaining an outcome.
Offline models usually admit better solutions than online models; this is the case in, e.g., temporal fair allocation settings \cite{bampis2018fairtime,igarashi2023repeated}.

\section{Sequentiality}
Another element of our framework is sequentiality of the problem instance.
This emphasizes the importance of \emph{order} over the timesteps.
The order matters when agents' have adaptive preferences (refer to the discussion of dynamic preferences in the Temporal Evolution subsection under Preferences), or where the contiguity of timesteps is important, as well as in online settings. For instance, in the context of AAAI executive council elections, a candidate agrees to serve on the council for three consecutive timesteps (and cannot participate immediately after completing their term); contiguity also matters for sequential committee elections~\cite{bredereck2020successivecommittee,bredereck2022committeechange,deltl2023seqcommittee}.
In contrast, in the scheduling model of \citet{elkind2022temporalslot} or \citet{bulteau2021jrperpetual}, timeslots can be rearranged arbitrarily, i.e., this model is fundamentally non-sequential.
While non-sequential settings have interesting computational problems in their own right \cite{boehmer2021bluesky}, sequential settings pose additional challenges.

\section{Solution Concepts} \label{sec:solutionconcepts}
The final element of our framework is the definition of suitable (temporal) solution concepts. 
For instance, one could consider the temporal extensions of popular fairness notions such as proportionality \cite{freeman2018dynamicproportional}, equitability \cite{elkind2022temporalslot}, or justified representation \cite{bulteau2021jrperpetual,chandak2023sequential}.
One may also wish to use time as a tool for allowing agents to be \emph{represented} eventually (global), consistently (local), or periodically (frequency-based).
We classify these approaches into three (non-exhaustive) broad categories.

\subsubsection*{\textbf{Global solution concepts}}
Global solution concepts evaluate agents' welfare across the \emph{entire} time horizon.
This means emphasizing the eventual outcome with respect to the welfare goal, even if it means that within certain timesteps the treatment of agents can be highly unequal.
An example of this approach is taking a global utilitarian or egalitarian view \citep{elkind2022temporalslot}, or achieving a pre-defined notion of fairness and welfare in offline models of scheduling~\cite{patro2022virtualconf}.

\subsubsection*{\textbf{Local solution concepts}}
Local solution concepts evaluate agents' welfare at specified time intervals.
This means emphasizing the welfare of agents either at each timestep, or for each pre-specified number of timesteps.
For instance, in the perpetual voting model~\cite{lackner2020perpetual}, decisions are made at each round, and welfare properties are considered for each agent up to and including the current round. 
In other models, such as sequential committee elections \cite{bredereck2020successivecommittee,bredereck2022committeechange,deltl2023seqcommittee}, the \emph{quality} of a committee may be defined and maintained as a goal at each timestep.

\subsubsection*{\textbf{Frequency-based solution concepts}}
Another possibility is to consider frequency-based properties, which can be formulated in terms of bounds on the \emph{number} of timesteps (be it consecutive or not) that have elapsed since the welfare of an agent was last addressed.
For instance, one could mandate that there should not be $\kappa$ timesteps that have elapsed such that an agent has not received a utility of at least $\gamma$. This idea was briefly mentioned by \citet{boehmer2021bluesky} in a different, more abstract social choice scenario. 
Several other problems arise as well: e.g., one could ask what combinations
of $\gamma$ and $\kappa$ can be accomplished by a specific voting rule (in the worst case, or on a specific instance), or by all
rules satisfying a given set of axioms; one could also ask what is the minimum attainable $\gamma$ for some fixed~$\kappa$.

\section{Research Directions/Challenges} \label{sec:directions}
We have suggested several dimensions according to which temporal multiwinner voting scenarios can be classified. Some of the ``points'' in the resulting multidimensional space have been considered already, but others remain unexplored and present opportunities for future work. On top of that, we will now highlight several interesting and challenging, yet fundamental directions for this field moving forward.
\subsubsection*{\textbf{Formalism of Solution Concepts and Goals}}
The study of fairness across time opens up opportunities for formulating novel solution concepts, either ones that are specific to the temporal setting, or generalized (temporal) variants of concepts developed for the traditional multiwinner election model.
For instance, popular concepts of \emph{representation} in the multiwinner election setting (e.g., justified representation and its variants \cite{aziz2017jr,sanchez2017proportional,peters2021fjr}) can be extended (\citet{bulteau2021jrperpetual} and \citet{chandak2023sequential} made first steps in this direction) and the associated computational problems can be studied.
Other welfare goals in multiwinner elections such as diversity \cite{brederek2018diversity,celis2018mwvfair,relia2022diversity} can similarly be explored.
Another direction that one could pursue would be considering a generalization of existing multiwinner voting rules to the temporal setting (this line of work was initiated by \citet{lackner2020perpetual} and \citet{lackner2023proportionalPV}), and investigating whether these rules satisfy the corresponding generalized temporal axioms (e.g., representation as discussed above).

\subsubsection*{\textbf{The Temporal Dimension}}
In the multiwinner voting literature, it is common to consider structural constraints
on voters' preferences, as a means to circumvent impossibility and 
computational hardness results \cite{BetzlerSU13,ElkindL15}. The temporal setting
may benefit from this approach, too. In particular, it may be of interest
to investigate the impact of structural constraints that are specific to the temporal setting.
For instance, when considering evolving preferences, one can place additional restrictions on how agents' preferences may change over time: e.g., perhaps agents' approval sets can only expand (as they learn about benefits of projects they were previously not aware of), or, alternatively, the preferences cannot change too drastically between two consecutive time periods; similarly, the set of available
candidates may evolve in a predictable manner, i.e., perhaps each candidate
is only present for a number of consecutive steps. Such constraints on candidates' availabilities and agents' 
temporal preferences can be viewed as a novel
structured preference domain~\cite{elp-survey2022}, and can therefore offer a pathway towards positive algorithmic and axiomatic results. 

Furthermore, we can consider settings where agents assign different importance to different timesteps (on top of preferences over candidates): e.g., in case of public projects, an agent may plan to be overseas during a specific time period and is therefore less interested in projects implemented during that period.
Allowing more expressive ballots will allow solutions that provide better welfare guarantees, but is likely to introduce a whole new set of computational challenges that must be dealt with.

Moreover, it is interesting to explore extensions of normative analysis into the temporal realm.
One could consider fairness criteria evaluated according to the worst-case, average-case or best-case with respect to timesteps, possibly with a discount factor. 
More broadly, when defining temporal extensions of notions of representation, rather than considering a simple additive variant across timesteps, one could define novel concepts that take into account the timesteps themselves in the definition.

\subsubsection*{\textbf{Does Time Hurt or Heal?}}
Another important question is to understand the effect of time in comparison to traditional single-shot multiwinner election models.
Apart from the impact (positive or negative) that it can have on agents' welfare (as discussed in previous sections), one can approach this question from a computational perspective:
does time make computing certain solution concepts easier (i.e., more algorithmically efficient), or does it add a computational hurdle?

For instance, in single-shot elections, the property of proportional justified representation~\cite{sanchez2017proportional} can only be accomplished by fairly sophisticated algorithms~\cite{brill2017phragmen}, whereas in perpetual voting with static preferences it is provided by a simple greedy algorithm~\cite{bulteau2021jrperpetual}. It would be interesting to explore whether similar results can be obtained for other proportionality and fairness concepts.

\subsubsection*{\textbf{Beyond Multiwinner Voting: Participatory Budgeting}}
Throughout the paper, we focused on the setting where the number of candidates
to be selected is specified as part of the input. One can also consider
a more general setting of {\em participatory budgeting}, where candidates
(projects) may have distinct costs, and there is a budget, so that the total cost
of the selected projects must remain within the budget \cite{aziz2021participatory}. While some of the work
we discussed considers this more expressive setting \cite{lackner2021longtermpb}, most of the literature 
focuses exclusively on the basic multiwinner scenario; it would be interesting
to see to what extent the existing positive results carry over to the richer participatory
budgeting domain.

\section{Conclusion}
The multiwinner voting literature has burgeoned in recent years, and significant progress has been made towards designing robust mechanisms for a fairer and more just society. 
The inclusion of temporal considerations introduces a plethora of novel research directions, which should be explored systematically and in-depth, with a focus on real-life applications.

We note that, while we aimed to be comprehensive in our analysis 
and systematization of the prior work, and proposed some 
new research directions to be investigated, there may be further dimensions
worth exploring. Overall, voting over time is a topic that is both practically important
and theoretically challenging, as we hope that our paper will stimulate
further research in this domain.

\section{Acknowledgments}
This work was supported by the AI Programme of The Alan Turing Institute.

\bibliography{abb,aaai24}

\end{document}